\documentstyle[twoside,fleqn,espcrc2]{article}

\newcommand{\be}{\begin{equation}}
\newcommand{\ee}{\end{equation}}
\newcommand{\bea}{\begin{eqnarray}}
\newcommand{\eea}{\end{eqnarray}}

\newcommand{\nn}{\nonumber}

\newcommand{\cO}{{\cal O}}

\newcommand{\cA}{{\cal A}}

\newcommand{\cT}{{\cal T}}

\newcommand{\cK}{{\cal K}}
\newcommand{\CF}{C_{\rm F}}

\newcommand{\alpham}{\alpha(\mu^2)}

\newcommand{\alphaq}{\alpha(Q^2)}
\newcommand{\al}{\alpha}
\newcommand{\la}{\lambda}
\newcommand{\m}{\mu}
\newcommand{\n}{\nu}
\newcommand{\g}{\gamma}

\newcommand{\AmS}{{\protect\the\textfont2
  A\kern-.1667em\lower.5ex\hbox{M}\kern-.125emS}}

\newcommand{\NPB}[3]{Nucl.\ Phys.\ {\bf B{#1}} (19{#2}) {#3}}
\newcommand{\PRD}[3]{Phys.\ Rev.\ {\bf D{#1}} (19{#2}) {#3}}
\newcommand{\PLB}[3]{Phys.\ Lett.\ {\bf B{#1}} (19{#2}) {#3}}
\newcommand{\PRL}[3]{Phys.\ Rev.\ Lett.\ {\bf {#1}} (19{#2}) {#3}}

\title{UV renormalons in QCD and their phenomenological implications
.\thanks{UAB-FT-424/97}}

\author{S. Peris\address{Grup de Fisica Teorica and IFAE, 
                 Universitat Autonoma de Barcelona, 
                 08193 Bellaterra (Barcelona), Spain. E-mail: peris@ifae.es}
        \thanks{Talk given at the workshop ``QCD 97", Montpellier,
France, July 1997. To appear in the proceedings.}
          \thanks{Work partially supported by research project
                CICYT-AEN95-0882.}
}
       
\begin{document}

\begin{abstract}

 
I report on some recent work done in collaboration with E. de Rafael on the
connection between ultraviolet renormalons in QCD and Nambu-Jona-Lasinio-like
Lagrangians as its effective description at low energies. 

\end{abstract}

\maketitle
 
In a superconductor \cite{Ziman}, let us denote by $V$ the lowest order 
amplitude
for two electrons to scatter off each other through the exchange of a phonon
($V$ is a negative number). We shall assume both electrons are outside but
very close to the Fermi sphere and we shall approximate $V$ to be a 
constant. One
can see that the total amplitude for this scattering process is actually an
infinite sum
\be
\label{one}
T\approx V\ \left( 1 + \g V + \g^2 V^2 + ...\right)
\ee
where $\g \sim \kappa^{-1} \log(E/\omega)$, with $\kappa >0 $ a constant 
depending on the metal, $\omega$ a 
parameter describing the effective interaction range (in momentum 
space) and $E$ is
the total energy of the electron pair. Under normal conditions $\g V$ is
small and the above series can be truncated after the firstfew terms.
However, when $E$ is very small more and more terms become important and one
has to consider the resummation of the series (\ref{one}):
\be
\label{two}
T = {V\over 1 - \g V}\ \ .
\ee
This matrix element has a pole at 
\be
\label{three}
E= \omega e^{- \kappa /\vert V \vert } \ ,
\ee
clearly something goes wrong with our perturbative description when the
energy reaches the value of eq. (\ref{three}). Of course what goes wrong is
that Cooper pairs form and the phenomenon of superconductivity takes place.
The dependence of the energy on the strength of the interaction (i.e. $V$)
takes a {\it nonanalytic} form such as eq. (\ref{three}). 
This is not surprising
physically, one should not have expected to be able to describe a
nonperturbative effect like the formation of Cooper pairs just by
analytically continuing in the strength of the interaction $V$ starting from
the perturbative regime (eqs. (\ref{one}-\ref{three})). This is an example of 
how perturbation theory may signal the existence of nonperturbative effects.

With an ordinary superconductor this is of course not the end of the story.
One can go a long way beyond eqs. (\ref{two},\ref{three}) by applying the
machinery of the Bogoliubov transformation to actually solve the 
quantum-mechanical system, describe its excitations and so on. The reason is
of course 
because the phenomenon of superconductivity is a rather well-understood one.

How about QCD ? Regretfully it is very clear that QCD is much more difficult
as a nonperturbative theory. But perhaps the previous example may make us
harbor some hope that studying the analytic properties of the perturbative
series may shed some light on some aspects of the nonperturbative
dynamics. Therefore, in this sense, one may ask: 

\rightline{}

{\it are there specific perturbative signatures of chiral symmetry breaking
  ?} \cite{Wilczek}

\rightline{}

To try to answer this question, let us adopt one simplification right from 
the start and take the large-$N_c$
limit of massless QCD \cite{Nc}. This limit 
presumably captures all the important
ingredients of the full QCD and in particular allows a proof of chiral
symmetry breaking \cite{Coleman-Witten}. 

We shall study a concrete example: let 
$\cA (Q^2)$ be the Adler function defined by ($Q^2\equiv -q^2$, with $Q^2>0$
for $q^2$--spacelike)
\be 
\label{four} {\cal A}(Q^2) \equiv -Q^2\frac{\partial \Pi}{\partial 
Q^2}\,.
\ee
where
\bea
\label{five}
\Pi^{\mu\nu}(q) & = &  i\int\, d^4 x e^{iq\cdot x}\langle 0\mid
 \mbox{\rm T}\left\{V^{\mu} (x)V^{\nu}(0)\right\}\mid 0\rangle \nn \\
 & = & -\left(g^{\mu\nu} q^2-q^{\mu}q^{\nu}\right)\Pi(q^2)
\eea
with 
\be  V^{\mu}=\frac{1}{2} \ 
(:\bar{u}\gamma^{\mu}u:-:\bar{d}\gamma^{\mu}d:) 
\ee
the vector-isovector quark current. Within the context of QED, it was
 recognized long ago by 't Hooft \cite{'t Hooft} and Lautrup \cite{Lautrup} 
that the exchange of
 the effective charge, i.e. an infinite
 string of ``vacuum polarization insertions'' 
\cite{bubble} 
 is responsible for an $n$! behavior of the $n$-th coefficient in the
 perturbative expansion in the coupling constant when $n$ is very large. A
 series of this sort has zero radius of convergence and as such can not be
 resummed (i.e. analytically continued in the coupling constant).

For a nonabelian theory like QCD the explicit construction of the
corresponding effective charge is of course much more difficult. However it 
can be worked
out \cite{Watson} with the use of what is called the pinch technique
\cite{pinch} with properties completely
analogous to those of the QED one\cite{deR}. 
In particular one also reproduces the full
first coefficient of the $\beta$ function. This is an important ingredient when
trying to understand chiral symmetry breaking since the $\beta$ function is the
only seed of scale symmetry breaking in perturbation theory and, clearly,
chiral symmetry breaking is intimately linked to the 
appearance of scales in QCD.  

In the particular case of the Adler function in QCD 
suffice it to say that
when the euclidean virtual momentum going through the internal gluonic bubble
chain $k_E$ is much smaller than the large external momentum $Q$ one can
approximate:
\bea
\label{seven}
\cA(Q^2)\vert_{\mbox{\small{\rm IR}}}
& \approx &\frac{N_c}{16\pi^2}\,C_{F}\,\frac{1}{2\pi b_0} \nn \\
& &\int_{0}^{\infty}dw\, e^{-\frac{w}{b_0\alphaq}}\:\frac{w}{2-w} \,
\eea
where the change of variables 
\be 
\label{sevenprime}
w/2=- b_0\alphaq \log k_E^2/Q^2\,,
\ee
has been made. Here $b_0 > 0$ is the first coefficient of the QCD $\beta$
function. Equation (\ref{seven}) is already written in the form of a Borel
transform \cite{Zinn-Justin}. One can obtain the perturbative series in 
$\alphaq$
from (\ref{seven}) by expanding the fraction in powers of $w$. Upon
integration this produces an $n$! growth of the perturbative coefficients that
yields a divergent series. The fraction $w/2-w$ in eq. (\ref{seven}) 
is obtained after analytically continuing in the $w$ plane for $w$
complex. This can be considered as a resummation of the perturbative series in
$\alphaq$ insofar as the integral exists.

But the integral in eq. (\ref{seven}) does not exist. The integrand has a pole
at $w=2$. This pole is called an infrared renormalon and it can
  be traced back to a singularity at the Landau pole in momentum space, 
$\Lambda_L$ \cite{beta}. We are in a similar situation as in
  eq. (\ref{two}). If perturbation theory were the whole story
  eq. (\ref{seven}) would imply that $\cA(Q^2)$ is ambiguous, a most
  disastrous conclusion for a physical observable ! Of course perturbation
  theory is not the whole story and it is only when nonperturbative
  contributions are added to the perturbative
  contributions to eq. (\ref{four}) that the final answer for $\cA(Q^2)$ is
  physical and unambiguous \cite{Mueller}. But just as we did in our initial
  example of the superconducting metal we can use perturbation theory for
  guessing what sort of nonperturbative contributions we are missing in
  eq. (\ref{seven}) by parametrizing its ambiguity. This can be done  
simply by choosing the
  integration contour from above or from below the singularity at $w=2$, 
\be
\label{eight}
\delta\cA(Q^2)\vert_{\mbox{\small{\rm IR}}}\sim e^{-2/b_0 \alpha(Q)} 
\sim \left(\frac{\Lambda_L}{Q}\right)^4\,,
\ee
where the one-loop running coupling constant has been used in the last
step\cite{twobeta}. Equation (\ref{eight}) suggests the
existence of a nonperturbative contribution behaving like $Q^{-4}$ at large
$Q^2$. This is how perturbation theory hints at the gluon condensate.

Let us go back to the Adler function and consider now the contribution in 
the regime where $k_E^2 >> Q^2$. One finds 
\bea
\label{nine}
\cA (Q^2)\vert_{\mbox{\small{\rm UV}}} & \approx &
\frac{N_c}{16\pi^2}\,\CF
\,\frac{4}{9}\,\frac{1}{2\pi b_0} \nn \\
& & \int_{0}^{\infty}dw\, e^{-\frac{w}{b_0\alphaq}}\ \frac{1}{(1+w)^2}\ 
\eea
 with $w=b_0 \alphaq \log k_E^2/Q^2$. The singularity in the $w$ plane is now
 at $w=-1$ and
 consequently the integral can be carried out. This singularity is called an
 ultraviolet renormalon. 
Equation (\ref{nine}) could
 then be considered a resummation of the perturbative series that is obtained
 by expanding the fraction in powers of the variable $w$. This produces the
 alternating 
series $\sim N_c (-1)^n (b_0\alphaq)^n n$! . Unlike the case of the infrared
 renormalon, perturbation theory does not send any signal about the existence
 of nonperturbative contributions in the form of an ambiguity caused by a
 singularity in the integration region\cite{Qtwo}.

However, as Vainshtein and Zakharov \cite{V-Z} first noticed, 
the contribution from two bubble chains\cite{fig}
actually dominates the asymptotics of the perturbative
coefficients in $\alpha$ over the contribution coming from just one chain,
eq. (\ref{nine}). The way to understand this is as follows: the vacuum
polarization tensor (\ref{five}) verifies 
\bea
\label{ten}
& &\Pi^{\m \n}(q)\ v^\m v^\n = \nn \\
\frac{-i}{2} \int_{k_E^2 \ge Q^2}
\frac{d^4k}{(2\pi)^4}\ & &\frac{4 \pi \alpha(k_E^2)}{k^2}\ \langle v\mid
\frac{\cT}{k^2} \mid v
\rangle\, \ 
\eea
where $v$ is taken as an external field coupled to the isospin current and  
\bea
\label{eleven}
\cT \equiv i\ k^2 & \int d^4x &e^{ik\cdot x}\ \mbox{\rm T}\Bigg\{:\bar
q(x)\gamma^\m
\frac{\la^a}{2}q(x):\ \nn \\
& &:\bar q(0)\gamma_\m \frac{\la^a}{2}q(0):\Bigg\}\, ,
\eea 
where $\la^a$ are the Gell-Mann matrices in color space. 
The crucial observation of V-Z 
is that $\cT$ admits an operator product expansion at large $k^2$ 
whose first physically relevant terms start at {\it dimension six} 
($F_{\m \n}(x)^{ext}$ below is the field strength of the field $v$):
\bea
\label{twelve}
\cT & \approx & N_c \frac{c_1(k^2)}{k^2}\ \bar q(0)\ 
\gamma^\m \ D^\al F_{\al \m}^{ext}(0) \ q(0) \nn \\
& + & \frac{c_2^V(k^2)}{k^2} \left[
\sum_{a,b} \mid \bar q^a_L \gamma^\m q^b_L\mid^2 + (\mbox{\rm L} \rightarrow
\mbox{\rm R})\right] \nn \\
& + & \frac{c_2^S(k^2)}{k^2} \left[ \sum_{a,b} (\bar q^a_L q^b_R) \
(\bar q^b_R q^a_L) \right] \nn \\
& + & N_c \frac{c_3(k^2)}{k^2} 
\ f_{ABC} G_{\m\n}^A G^{\n B}_{\rho} G^{\rho\m C} +  ...  \ ,
\eea
where the ellipses stand for dimension-eight operators and higher. The
contribution of one chain yields $c_1=2/3, c_2^V=c_2^S=c_3=0$. At the
level of two chains, one has 
$c_1=0, c_2^V= 2 N_c \pi \al(k^2) , c_2^S= - 32 
N_c \pi \al(k^2), c_3=0$\cite{PdeR}\cite{Maxwell}. 
As one considers diagrams with chains of an
increasing complexity one will find that the coefficients $c_i$ in the OPE
have themselves an expansion in powers of $\al(k^2)$.

Computing the matrix element $\langle v\mid \cT \mid v \rangle$ in
eq. (\ref{ten}) to get to $\cA(Q^2)$ in eq. (\ref{four}) 
is tantamount to computing the mixing of the operators
(\ref{twelve}) into the operator $(\partial_{\m} F_{ext}^{\m \n})^2$. We are
only interested in this mixing at one loop since this order resums equal
powers of the logarithms and the coupling constant $\alphaq$, i.e. terms of
the form $\left[\alphaq \log k^2/Q^2 \right]^n$ which are leading
contributions to the asymptotics $\sim \al^n n$! since every power of $\log$
is eventually transformed into a power of $n$. Considering the
mixing only at one loop, the triple gluon operator $f_{ABC} G_{\m\n}^A
G^{\n B}_{\rho} G^{\rho\m C} $ drops out of the problem since it does not mix
with the operators of eq. (\ref{twelve}) nor with $(\partial_{\m} F_{ext}^{\m
  \n})^2$ \cite{Narison-Tarrach}. For simplicity of 
presentation I will also ignore
the presence of the anomalous dimension matrix. This will change nothing of
the physics I am discussing 
and, if need be, it can always be taken into account
\cite{Beneke-Braun-Kivel}. 

The final contribution to the Adler function can be calculated 
through eq. (\ref{ten}) and turns out to be dominated by {\it the vector-like 
four-quark operator}, 
whose coefficient is $c_2^V$ in eq. (\ref{twelve}), and yields
\bea
\label{thirteen}
\cA(Q^2)\mid^{\mbox{\rm \small two\ chains}}_{\mbox{\rm \small UV}}&=&
 \frac{2}{9} \frac{N_c}{16\pi^2}\
\left(\frac{N_c}{2}\,\frac{1}{2\pi b_0}\right)^2 \nn \\
\int_{0}^\infty dw \ e^{- \frac{w}{b_0\alphaq}}& &\frac{w}{(1+w)^3} \ \ ,\ 
\eea 
which in perturbation theory means an asymptotic behavior $\sim N_c (-1)^n
(b_0 \alphaq)^n (n+1)$! . Diagrams with an increasing
complexity multiply the result (\ref{thirteen}) by a factor $\chi$
whose value is beyond any known approximation scheme, presumably involving an
infinite-number-of-loops calculation. I would like to stress here that the fact
that a four-quark operator is the one that dominates is true 
not only for the vector channel and the Adler function but also for the
scalar channel and its corresponding associated function (and also for
the axial-vector and pseudoscalar respectively)\cite{PdeR}. 

It is of course natural to try to associate this four-quark operators with some
sort of an effective Nambu-Jona-Lasinio (NJL) Lagrangian \cite{NJL}
\cite{V-Z}. However two obvious difficulties are immediately encountered. 
Firstly a NJL-like Lagrangian, being a Lagrangian, 
has to be local which in turn
requires a scale to appear in the denominator of the dimension-six four-quark
operators of eq. (\ref{twelve}) instead of the floating virtual momentum
$k^2$. Secondly, a NJL type of  Lagrangian can make sense only as an effective
description at low energies, whereas up to now all momenta $k^2$ and $Q^2$
have been considered to be large. As we shall see the 
above two points are actually related. 

What does this mean ?  
Let us look at the same calculations but now in the low-energy regime,
i.e. when $Q^2\to 0$. The contribution to the Adler function from the diagram
with two chains is then
\bea
\label{fourteen}
\cA(Q^2)&\sim & N_c^3 \ Q^2 \nn \\
& &\int_{Q^2}^\infty
\frac{dk^2_E}{k_E^4}\ \left[\alpha(k_E^2)\right]^2\ \log^2\frac{k^2_E}{Q^2} \ .
\eea

Using the previous change of variable $w=b_0 \alphaq \log k_E^2/Q^2$ where now
$\alphaq$ is $defined$ as\cite{anal}
\be
\label{fifteen}
\alphaq\equiv \frac{\alpham}{ \Big(1 + b_0 \alpham \log(Q^2/\mu^2)\Big)} 
\ee
for the entire range of its argument $Q^2$ (so that for instance $\alphaq < 0$
if $Q^2 < \Lambda_L^2$) one finds
\bea
\label{sixteen}
\cA(Q^2)& \sim & \frac{N_c}{b_0\mid \alphaq\mid}\nn \\
& &\int_0^\infty  dw \ e^{- \frac{w}{b_0 \mid \alphaq\mid }}\ 
\frac{w^2}{(1 - w)^2}\ \ .
\eea

Unlike in the usual case ($Q^2 > \Lambda^2_L$), now ($Q^2 < \Lambda^2_L$) we
find a pole in the integration region for the contribution of the ultraviolet
renormalon. Contributions from more complex diagrams will introduce a
multiplicative factor in front of eq. (\ref{sixteen}). This singularity causes
the integral to be ambiguous. This ambiguity can be parametrized as
\bea
\label{seventeen}
& &\delta \cA(Q^2)_{Q^2<\Lambda_{L}^2} = 
- \ \cK \frac{N_c}{16\pi^2}\
\left(\frac{N_c}{2}\, \frac{1}{2\pi b_0}\right)^2\nn \\
 & & \quad \quad \frac{Q^2}{\Lambda_L^2}\ 
\left(\frac{1}{9} 
\log^2\frac{\Lambda_L^2}{Q^2} + ...\right) \ ,
\eea
with $\cK$ an unknown constant. This ambiguity 
is of the same form as the insertion of the dimension-six operator
\bea
\label{eighteen}
& &\cO_V = - \frac{8\pi^2 G_V}{N_c \Lambda_{\chi}^2}\nn \\
& &\sum_{\mbox{\rm \small a,b=flavour}}  
\left[ (\bar q^a_L \gamma^\m q^b_L)\ (\bar q^b_L \gamma_\m q^a_L) 
+ (\mbox{\rm L} \rightarrow \mbox{\rm R}) \right]\qquad 
\eea
provided one interprets $\Lambda_L^2$ as the momentum cutoff in the loops, 
$\Lambda_{\chi} \simeq \Lambda_L$ and one identifies 
$4 G_V \equiv - \cK\ \left(\frac{1}{2\pi b_{0}}\frac{N_c}{2}\right)^2$. 
This is precisely one of the four-quark
operators appearing in the extended Nambu-Jona-Lasinio model \cite{ENJL}
. Repeating the same analysis for the pseudoscalar two-point
function will lead to the other four-quark operator of this model. These two
are the four-quark operators we already encountered in 
eq. (\ref{twelve}) but now they
are truly local operators, i.e. they have a true constant in the denominator,
not the virtual floating momentum $k^2$ as in
eq. (\ref{twelve}). Therefore they can be considered as an honest to goodness
interaction in an effective local Lagrangian.

The ENJL model has proven itself very successful in predicting the low-energy
chiral Lagrangian of Gasser and Leutwyler\cite{G-L} but its possible connection
with QCD has remained very elusive so far. Renormalons suggest a, in my
opinion, very remarkable connection between perturbation theory in QCD and
ENJL, and answer with a ``yes'' the question posed at the beginning since it
is 
the same four-quark operators that are responsible for i) the asymptotic
behavior of the perturbative series of two-point functions in QCD and ii) for
the breakdown of chiral symmetry. Although this does not constitute a proof,
it may be considered unlikely that it be a pure coincidence. 

Returning to my initial example, it is interesting how also superconductivity
is the effect of a four-fermion operator turning relevant at low
energies \cite{Polchinski}. At the very least I
think one can take these considerations as some intriguing evidence in favor
of ENJL as an effective description of QCD in the low-energy regime.

{\bf Acknowledgements}

I would like to thank F. Lopez Aguilar for discussions on Superconductivity.

{\bf Discussions}

{\bf A. Di Giacomo}, Pisa.

{\it In studying field correlators on the lattice (talk by M. D'Elia), the UV
  cutoff can be changed and moved towards the IR one by the ``cooling''
  technique. Can your analysis help in disentangling the contribution of
  renormalons to the gluon condensate by looking at the cutoff dependence 
  of the data ?}

{\bf S. Peris}

{\it Given my present understanding I don't quite see how. 
  Four-quark operators
  either in the form of eq. (13) or eq. (19) are always related to UV
  renormalons, i.e. to the regime where the virtual gluon momentum $k_E$ is
  much larger than the external momentum $Q$. However, the gluon condensate is
  related to IR renormalons, i.e. to the opposite situation $Q>>k_E$.}


\begin{thebibliography}{9}
\bibitem{beta} This is strictly true only with a one-loop $\beta$ function
               \cite{Peris-deR}.
\bibitem{twobeta} Without loss of generality one could consider a scheme where
  only the first two coefficients of the $\beta$ function are non-zero\cite{'t
  Hooft} with the same physical conclusions.
\bibitem{anal} This can be considered as an analytic continuation in
  $\alpha(\mu^2) \log Q^2/\mu^2$.
\end{thebibliography}
\end{document}